# Space observation on detoxing the unhealthy air quality during COVID-19 pandemic in India

Prabhat Kumar[1]*, Rohit Kumar Kasera[2], S Suresh[3]

[1, 3] Department of Computer Science, Institute of Science, Banaras Hindu University, Varanasi – 221 005, India.
[2] Department of Computer Science and Engineering, Assam University, Silchar, Assam – 788011, India
Email: prabhat.kumar13@bhu.ac.in

*Abstract:* The purpose of this study has extremely dedicated to expose the correlation between coronavirus pandemic and space observation on unhealthy air quality in India. The world has undergone in lockdown to break the chain of corona virus infection. The Air Quality Index (AQI) has started to improve after the commencement of lockdown due to industrial and transportation sectors temporally closed. This study compiled the data recently released by NASA (National Aeronautics and Space Administration), ESA (European Space Agency) and ISRO (Indian Space and Research Organization). In this paper, we have discussed about the space observation on Nitrogen Dioxide ($NO_2$), Aerosol Optical Depth (AOD), PM2.5, and PM10 influenced the air quality across the various region of India. We analyzed the detoxing of air quality before and during the lockdown period over the same time frame of current and the previous year. The result has shown the positive impact on the detoxing of unhealthy air quality during lockdown stated as the emission of $NO_2$ has reduced to 40% - 50% and optical level of aerosol indexed at low compared to the last 20 years in northern India.

## I. INTRODUCTION

This The World Health Organization (WHO) has declared the COVID-19 as a global pandemic across the world, was the first case acknowledged in the Wuhan city, China in late December 2019 (Chen et al., 2020; WHO, 2020). The human to human transmission properties of coronavirus becomes a critical situation to control overspreading the infection. As of March 7, 2020, a total of 212 countries and territories around the world have got affected with 3,834,142 confirmed cases and 265,240 deaths (*Countries where Coronavirus has spread - Worldometer*, n.d.). The clinical characteristics of COVID-19 have initial symptoms included as fever, cough, dyspnea, myalgia or fatigue, sputum production, headache, hemoptysis, and diarrhea (Jiang et al., 2020). The fatality rate of COVID-19 in Italy and China is similar, approximately 2-3% (Porcheddu et al., 2020). The coronavirus has covered all the major cities of China very quickly and explored more than 20 countries on January 30, 2020. By the analysis of the COVID-19 influencing rates and death rates, the China government has declared the Wuhan city under quarantine on January 23, 2020 (Muhammad et al., 2020), and subsequently, the whole country also falls under lockdown to slow down the affection. The isolation, quarantine, social distancing, and community containment are a powerful way to break the chain of human to human transmission of coronavirus (Wilder-Smith & Freedman, 2020). The worldwide public health emergency was declared by WHO on January 30, 2020.

The Ministry of Health & Family Welfare of India has confirmed the COVID-19 pandemic was reported on January 30, 2020, in Kerala, India (PIB Delhi, 2020). Maharashtra is now the most affected state of COVID-19 confirmed cases and the second most affected state in Gujarat. As on June 07, 2020, 01:44 AM IST, the total confirmed 2,46,622 cases, 1,20,968 active cases, 1,18,695 recovered (including 1 migration) and 6,946 people deaths in the country (*Coronavirus Outbreak in India - covid19india.org*, 2020; *MoHFW | Home*, n.d.). The Indian government also committed to taking all necessary actions to outbreak the spreading of coronavirus. The public curfew for 14-hours was declared on March 22, 2020, when the number of positive cases crossed 500 in India. Approximately, 1.3 billion people of India imported a self-imposed quarantine for 14 hours to save the nation (*India coronavirus: Millions under lockdown as major cities restrict daily life - CNN*, n.d.). The prime minister of India announced the completely nation-wide lockdown for 21-days (25 March – 14 April) on March 24, 2020, to stay the 136 million people at home (*India, Day 1: World's Largest Coronavirus Lockdown Begins - The New York Times*, n.d.). After completion of lockdown 1.0, the Prime Minister Narendra Modi announced that India remains in lockdown till May 3, 2020, fall in lockdown 2.0 (*In coronavirus lockdown extension, Modi wields stick, offers carrot on exit route - Coronavirus Outbreak News*, n.d.). The Government of India (GoI) and Ministry of Home Affairs (MHA) have extended the period of lockdown for two weeks till May 17, 2020, announced on May 1, 2020 and lockdown 3.0 going on (*India Lockdown Extension News: Lockdown extended by 2 weeks, India split into red, green and orange zones - The Economic Times*, n.d.).

Due to the lockdown in the whole country, public transportation services (either road or air) and industrial services are finally stopped excluding emergency services such as medical. The health advisors are also recommended to the people to stay safe and stay at home. According to the CNN-report, the airplane passengers have dropped traveling up to 96% due to coronavirus pandemic (*Airlines and TSA report 96% drop in air travel - CNNPolitics*, n.d.).

The Ministry of Electronics & Information Technology, Government of India launched smartphone-based application "Aarogya Setu", available at Google Play Store (for Android phones) and Apple App Store (for ioS) (*Aarogya Setu Mobile*

*App | MyGov.in*, n.d.). This application continuously monitors someone tested as corona positive and quickly notifies the users. The tracking functionality is done through Bluetooth and GPS-based location services (*aarogya setu app: How to use Aarogya Setu app and find out if you have coronavirus symptoms - The Economic Times*, n.d.).

The contribution of our work is to assess the quality of air including the primary pollutants in the atmosphere during COVID-19 fatality in India. This work is accomplished with regards to NASA Earth Science Data (*Earthdata*, n.d.), Giovanni (*Giovanni*, n.d.), ISRO (Indian Space and Research Origination) and European Space Agency (*European Space Agency*, n.d.). The remaining sections of this paper are arranged as the second section contains related studies, Community Mobility Report mentioned in the third section, the assessment of air quality observation described in the fourth section and discussion and conclusion are in the fifth section.

## II. RELATED STUDIES

One of the famous English quotes stated by Billy Graham "When wealth is lost, nothing is lost; when health is lost, something is lost; when the character is lost, all is lost". The COVID-19 has made the nastiest crisis and horrific effect on human health and the world economy. However, the positive side of environmental pollution due to the limited exercise of transportation and industrial sectors (Dutheil et al., 2020). The transportation sector is an extremely impressive source of Nitrogen dioxide ($NO_2$) for environmental pollution (He et al., 2020). The emission of global carbon dioxide ($CO_2$) in the world (2020), has also reduced to 0.3% to 1.2% (*The Coronavirus Outbreak Is Curbing China's CO2 Emissions - IEEE Spectrum*, n.d.). Analysis of the consumption rate of coal and crude oil for two week period, the emission of $CO_2$ has reduced to 25% predominantly in China and 6% globally (*Analysis: Coronavirus temporarily reduced China's CO2 emissions by a quarter*, n.d.). The quality of air directly influences the human health and immune system, also responsible for respiratory diseases such as asthma, chronic obstructive pulmonary disease (COPD), pulmonary fibrosis, pneumonia, and lung cancer (Brauer, 2010). The burden of disease data from 1990 to 2015 has analyzed at various factors including global, regional, and country-based, we get approximately 4·2 million human deaths and 103·1 million human lost their healthy life in 2015 (Cohen et al., 2017). The Fine particulate matter (PM2.5) has responsible for 3.6 million humans life losses annually in European nations (Thematic Strategy on air pollution (Text with EEA relevance), n.d.). The United States Environmental Protection Agency (EPA) reported that 92% of the population of China experienced more 120-hour unhealthy air and 1.6 million human deaths per year covered 17% of all deaths (Rohde & Muller, 2015). The long-term emission of $NO_2$ is statistically proportional to increasing respiratory mortality (Beelen et al., 2008). The chemical reactions among primary pollutants in the atmosphere are responsible to generate the secondary pollutants i.e. very harmful for human health such as acid rain, ground-level ozone, etc. The conversion rate of $NO_2$ and Sulfur dioxide ($SO_2$) are significant because of the source of some harmful secondary pollutants nitric acid ($HNO_3$) and ozone ($O_3$) (Khoder, 2002). The WHO has intentionally advised that instance of polluted airs including $NO_2$, $SO_2$, $O_3$, and particulate matters are responsible for significantly affect human health (*Air pollution and health - Air Pollution - Environmental Policy - UNECE*, n.d.).

## III. COMMUNITY MOBILITY REPORT DURING COVID-19 PANDEMIC IN INDIA

Google has released the report on community mobility from March 21, 2020, to May 02, 2020 (*COVID-19 Community Mobility Reports*, n.d.). This comparison report is based on traveling and staying days at different places, compared with the baseline period for five weeks (January 3, 2020 – February 6, 2020). The reported community places are marked noted as Grocery & pharmacy, Parks, Transit stations, Retail & recreation, Residential, and Workplaces. The community mobility report on alphabetically order from the top five states and also in overall India is presented in figure 1 (*COVID-19 Community Mobility Reports*, n.d.).

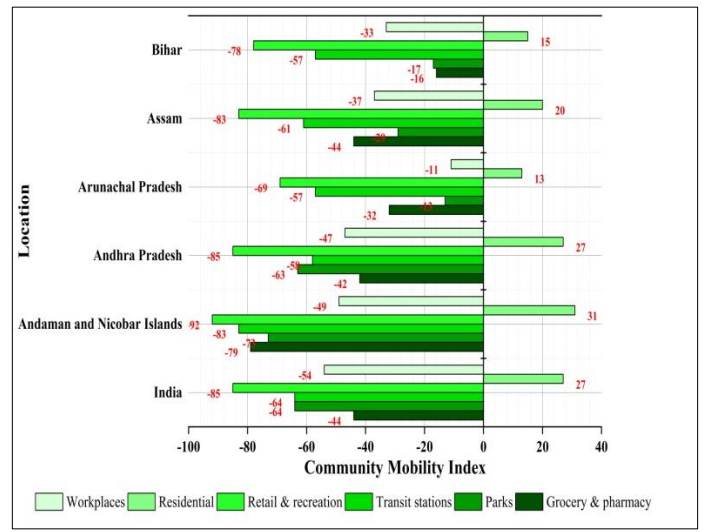

Fig. 1. Community mobility report for India based on Google tracking.

## IV. ASSESSING AIR QUALITY DURING THE LOCKDOWN IN INDIA

Air pollution is a worldwide problem for both developing and developed countries that affect human health. According to the most recent World Air Quality Report from IQAir, the ten most polluted cities in the world contain six Indian cities (*India dominates the list of the world's most polluted cities | World Economic Forum*, n.d.). The WHO estimates that approximately 4.2 million people are dying due to the worst outdoor air pollution (*Air pollution*, n.d.).

The ESA (European Space Agency), NASA (National Aeronautics and Space Administration) and Indian Space Research Organization (ISRO) released the considerable evidence regarding the improvement of air quality in various regions of India during the lockdown period using their weather satellites. These satellites produced various satellite images before and during the lockdown period. The contribution of listed space agencies monitoring the air quality while lockdown in India, described in the following sub-sections.

### A. Copernicus Sentinel-5P satellite

Recently, the lockdown not only assists to break the chain of spreading coronavirus but also responsible for the reduction of NO2 concentrations across the major cities of India. The major sources of NO2 in the atmosphere are industrial, transportation, and power plant sectors. The satellite image produced using data from Copernicus Sentinel-5P satellite, from the European Union Copernicus program, shown that decline of seeing level around 40-50% during nationwide lockdown (*ESA - Air pollution drops in India following lockdown*, n.d.). Figure 1 demonstrates the comparison of NO2 concentrations between currently specified periods and at the same time frame of last year. The current period dated as before lockdown (January 01, 2020 – March 24, 2020) and during lockdown (March 25, 2020 – April 20, 2020) compared with last year same period dated as (January 01, 2019 – March 24, 2019) and (March 25, 2019 – April 20, 2019) respectively. The dark red color shows the highest emission and white color marks show a low emission of NO2 in figure 2.

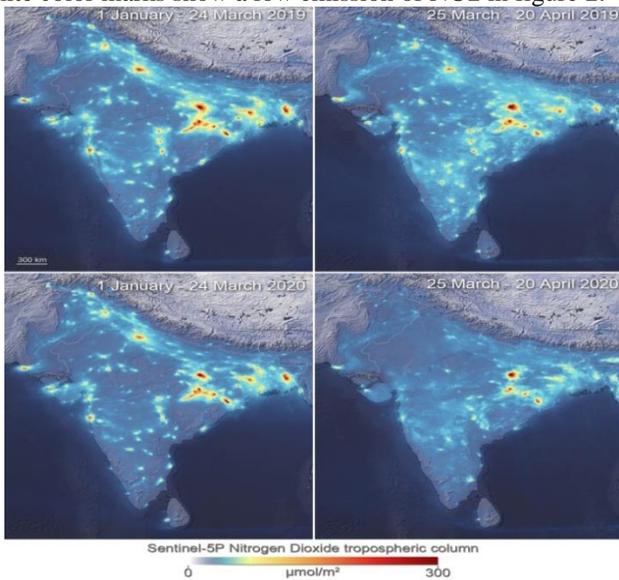

Fig. 2. Nitrogen Dioxide (NO$_2$) concentrations over India (*ESA 2020*, n.d.).

We observed the negligible dissimilarity of emission before the lockdown mentioned period compared with last year at the same time frame. During the lockdown, the specified period got interesting reducible up to 40-50% emission rate of NO2 as released in last year. After the announcement of nationwide lockdown on March 25, 2020, 136 million people resident at their home and avoid the use of non-essential services like as shops, markets and worship places were closed and only non-avoidable services such as water supply, electricity, medicals are remaining active. Due to avoiding all types of non-essential services are played the major roles in the reduction of NO2 in the atmosphere

### B. NASA's Terra satellite

The NASA satellite produced the level of aerosol in northern India, using data retrieved from Moderate Resolution Imaging Spectroradiometer (MODIS) (*MODIS Web*, n.d.). The aerosol level observed significantly low as compared to the last 20 years record in northern India (*Airborne Particle Levels Plummet in Northern India*, n.d.).

The industrial and transportation sectors are also responsible for increasing the aerosol level in the atmosphere. The aerosol is available in the form of liquid and solid and reduces the visibility on the earth. This also makes a harmful effect on the human lungs and hearts. The depth of aerosol does not exist in hazy condition until its thickness occurs less than 0.1, but when crossed 1 or more, the hazy condition arise. The AOD measures the prevention rate of reaching the sunlight to the earth's surface. The aerosol particles in the atmosphere help to block or scattered the sunlight (*ESRL Global Monitoring Laboratory - Global Radiation and Aerosols*, n.d.). The observation of Aerosol Optical Depth (AOD) is shown in figure 3 over the last five years (2016-2020).

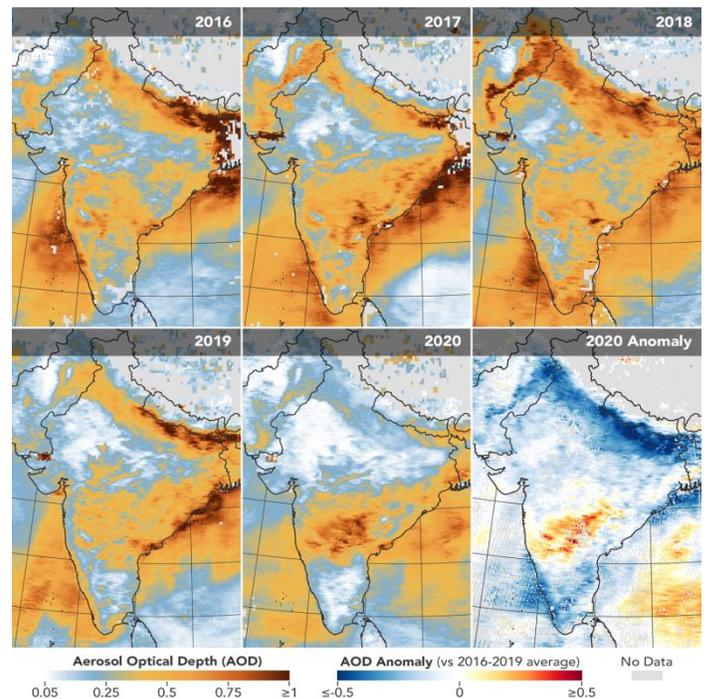

Fig. 3. Aerosol Optical Depth (AOD) level in Northern India(AOD in India (2016-2020), n.d.).

The data has captured during the same periods from March 31, 2020, to April 5, 2020, for each year of 2016-2020. In figure 2, the first five maps have shown the spreading of AOD in India over the years of 2016 to 2020. The last map shows AOD in 2020 of specified periods and compared with an average depth of aerosol from 2016 to 2019. This is observed that AOD goes down as compared to the average depth rates of the last five years.

### C. INSAT-3D

The ISRO launched INSAT-3D, an advanced weather forecasting satellite that monitors the humidity and temperature of the atmosphere for disaster management (*INSAT-3D - ISRO*, n.d.). The AOD level, PM2.5 and PM10 have measured before and during lockdown periods and satellite maps given in figure 4. The AOD including PM2.5 and PM10 has reduced in during lockdown period (March 25, 2020 – April 05, 2020) compared with before lockdown period (March 15, 2020- March 24, 2020). The analysis of reduced rates of aerosol levels has been dropped on the average of 28-40% overall country during lockdown period (Space based observation on changes in Air Quality during COVID-19 Lockdown period, n.d.). The air quality is being improved and the sky gets back to its original color i.e. bluer.

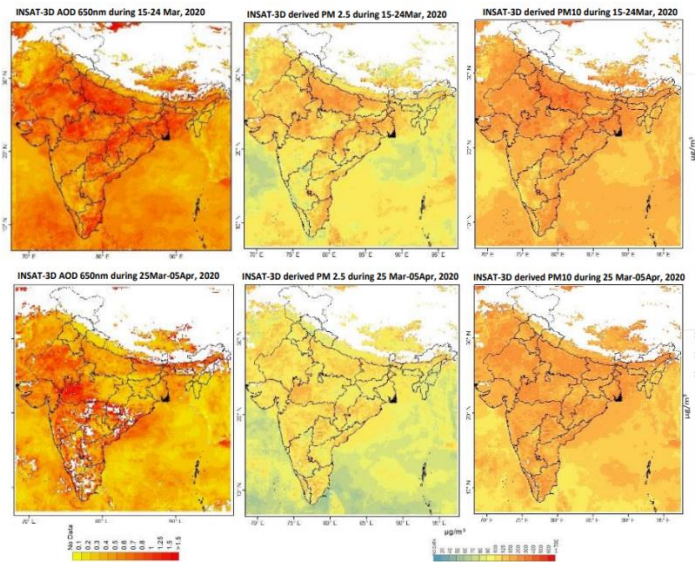

Fig. 4. AOD, PM2.5, and PM10 observation before and during the lockdown period in India.

### D. GIOVANNI – The Bridge between data and science

The NASA Goddard Earth Sciences (GES) Data and Information Services Center (DISC) launched the web-based application GIOVANNI (*Giovanni*, n.d.). This application provides an interactive and simple way to visualize and analyze the remote sensing data (*GES DISC Tools: Giovanni*, n.d.). The multi-disciplinary areas such as Aerosols, Atmospheric Chemistry, Atmospheric Dynamics, Cryosphere, Hydrology, Ocean Biology, Oceanography, and Water and Energy Cycle are integrated into the GIOVANNI platform. Various atmospheric affecting parameter values such as Aerosol Optical Depth, Air Temperature, $CO_2$, Incident Radiation, $NO_2$, $SO_2$, $SO_4$, Sea Surface Temperature, UV Exposure, etc. through the world-wide or particular areas can be easily visualized. The remote sensing data can be plotted in various ways such as Time Averaged Map, Monthly and Seasonal Averages, Scatter, Area Averaged (Static), Scatter, Time-Averaged (Interactive), Cross Section, Latitude-Pressure, Cross Section, Longitude-Pressure, Area-Averaged Differences, Area-Averaged, Zonal Mean, Histogram, etc. The starting and ending period can also be declared to extract the data and visualize the map for specified periods.

In figure 5, the first map presents the averaged time map of $NO_2$ spreading over India before the lockdown period (March 10, 2020 – March 24, 2020), and second last represents the spreading map during the lockdown period (March 25, 2020 – April 07, 2020).

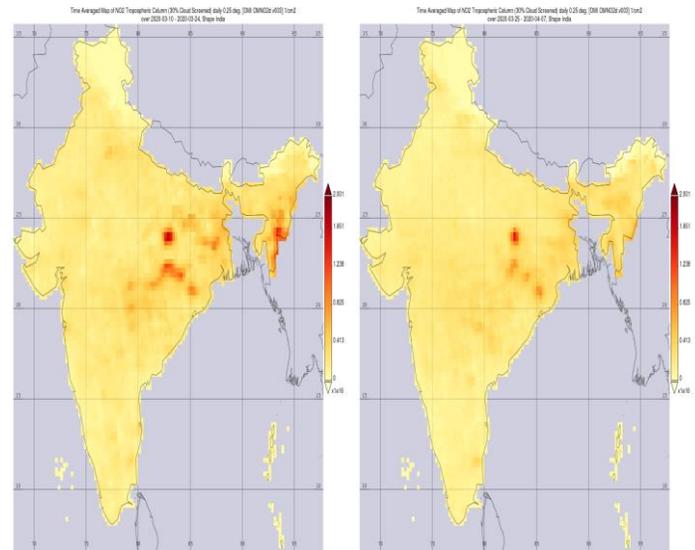

Fig. 5. Time-averaged map of $NO_2$ Tropospheric Column (30% Cloud Screened) over the before and during lockdown period.

The dark red dots represent the great emission of $NO_2$ over the specified area. The north-east region of India posses a high reduction of $NO_2$ as compared to other regions during the lockdown.

## V. DISCUSSION AND CONCLUSION

The COVID-19 has declared as a global pandemic for human beings but a blessing for the environment to regain itself where the pollution reduced. The positive impact on the environment may be temporary, but lockdown will be treated as one of the ways to control pollution. The space observation on emission rates of pollutant particles in the atmosphere is summarized in table 1. In this study, we analyzed the reduction of poor air quality in the atmosphere during the lockdown period as compared to before lockdown with the help of different weather and air quality observing satellites. We observed that some regions of India surely observed pure air quality as compared to the previous lockdown period. According to these observations, more future works should be conducted which covered additional factors. This study contained only atmosphere health observation during the lockdown period. We can also include various factors such as the affection rate of other diseases on the human being along with the pre-existence and post-existence of unhealthy pollutant particles due to COVID-19.

Table 1. Data acquisition of $NO_2$, AOD, PM2.5, and PM10 emission across India.

| Space Agency | Satellite | Period | Gases | Reduction (%) | Source |
|---|---|---|---|---|---|
| ESA | Sentinel-5P | (January 01- March 24, 2020) and (March 25- April 20, 2020) | $NO_2$ | 40% - 50% | (*ESA - Air pollution drops in India following lockdown*, n.d.) |
| NASA | Terra | March 31 – April 05, 2016 - 2020 | AOD | Lowest of last 20 years. | (*Airborne Particle Levels Plummet in Northern India*, |

| | | | | | n.d.) |
|---|---|---|---|---|---|
| ISRO | INSAT-3D | (March 15, 2020- March 24, 2020) and (March 25, 2020 – April 05, 2020) | AOD, PM2.5, PM10 | 28% - 40% | (*Space based observation on changes in Air Quality during COVID-19 Lockdown period*, n.d.) |
| NASA | GIOVANNI | (March 10, 2020 – March 24, 2020) and (March 25, 2020 – April 07, 2020) | NO$_2$ | Approx. 40% - 50% | (*Giovanni*, n.d.) |

## VI. ACKNOWLEDGEMENTS

The authors would like to thanks and regard to the ESA (European Space Agency), NASA (National Aeronautics and Space Administration) and ISRO (Indian Space and Research Origination).

REFERENCES

*aarogya setu app: How to use Aarogya Setu app and find out if you have coronavirus symptoms - The Economic Times*. (n.d.). Retrieved June 6, 2020, from https://economictimes.indiatimes.com/tech/software/how-to-use-aarogya-setu-app-and-find-out-if-you-have-covid-19-symptoms/articleshow/75023152.cms?from=mdr

*Aarogya Setu Mobile App | MyGov.in*. (n.d.). Retrieved June 6, 2020, from https://www.mygov.in/aarogya-setu-app/

*Air pollution*. (n.d.). Retrieved June 6, 2020, from https://www.who.int/health-topics/air-pollution#tab=tab_1

*Air pollution and health - Air Pollution - Environmental Policy - UNECE*. (n.d.). Retrieved June 6, 2020, from https://www.unece.org/environmental-policy/conventions/envlrtapwelcome/cross-sectoral-linkages/air-pollution-and-health.html

*Airborne Particle Levels Plummet in Northern India*. (n.d.). Retrieved June 6, 2020, from https://earthobservatory.nasa.gov/images/146596/airborne-particle-levels-plummet-in-northern-india

*Airlines and TSA report 96% drop in air travel - CNNPolitics*. (n.d.). Retrieved June 6, 2020, from https://edition.cnn.com/2020/04/09/politics/airline-passengers-decline/index.html

*Analysis: Coronavirus temporarily reduced China's CO2 emissions by a quarter*. (n.d.). Retrieved June 6, 2020, from https://www.carbonbrief.org/analysis-coronavirus-has-temporarily-reduced-chinas-co2-emissions-by-a-quarter

*AOD in India (2016-2020)*. (n.d.). Retrieved June 6, 2020, from https://eoimages.gsfc.nasa.gov/images/imagerecords/146000/146596/india_tmo_2016-2020_lrg.png

Beelen, R., Hoek, G., van den Brandt, P. A., Goldbohm, R. A., Fischer, P., Schouten, L. J., Jerrett, M., Hughes, E., Armstrong, B., & Brunekreef, B. (2008). Long-term effects of traffic-related air pollution on mortality in a Dutch cohort (NLCS-AIR study). *Environmental Health Perspectives*, *116*(2), 196–202. https://doi.org/10.1289/ehp.10767

Brauer, M. (2010). How much, how long, what, and where: Air pollution exposure assessment for epidemiologic studies of respiratory disease. *Proceedings of the American Thoracic Society*, *7*(2), 111–115. https://doi.org/10.1513/pats.200908-093RM

Chen, H., Guo, J., Wang, C., Luo, F., Yu, X., Zhang, W., Li, J., Zhao, D., Xu, D., Gong, Q., Liao, J., Yang, H., Hou, W., & Zhang, Y. (2020). Clinical characteristics and intrauterine vertical transmission potential of COVID-19 infection in nine pregnant women: a retrospective review of medical records. *The Lancet*, *395*(10226), 809–815. https://doi.org/10.1016/S0140-6736(20)30360-3

Cohen, A. J., Brauer, M., Burnett, R., Anderson, H. R., Frostad, J., Estep, K., Balakrishnan, K., Brunekreef, B., Dandona, L., Dandona, R., Feigin, V., Freedman, G., Hubbell, B., Jobling, A., Kan, H., Knibbs, L., Liu, Y., Martin, R., Morawska, L., … Forouzanfar, M. H. (2017). Estimates and 25-year trends of the global burden of disease attributable to ambient air pollution: an analysis of data from the Global Burden of Diseases Study 2015. *The Lancet*, *389*(10082), 1907–1918. https://doi.org/10.1016/S0140-6736(17)30505-6

*Coronavirus Outbreak in India - covid19india.org*. (2020). Covid19india. https://www.covid19india.org/

*Countries where Coronavirus has spread - Worldometer*. (n.d.). Retrieved June 6, 2020, from https://www.worldometers.info/coronavirus/countries-where-coronavirus-has-spread/

*COVID-19 Community Mobility Reports*. (n.d.). Retrieved June 6, 2020, from https://www.google.com/covid19/mobility/index.html?hl=en

Dutheil, F., Baker, J. S., & Navel, V. (2020). COVID-19 as a factor influencing air pollution? *Environmental Pollution*, *263*, 2019–2021. https://doi.org/10.1016/j.envpol.2020.114466

*Earthdata*. (n.d.). Retrieved June 6, 2020, from https://earthdata.nasa.gov/

*ESA - Air pollution drops in India following lockdown*. (n.d.). Retrieved June 6, 2020, from https://www.esa.int/Applications/Observing_the_Earth/Co

pernicus/Sentinel-5P/Air_pollution_drops_in_India_following_lockdown

*ESA 2020*. (n.d.). Retrieved May 12, 2020, from https://www.esa.int/var/esa/storage/images/esa_multimedia/images/2020/04/nitrogen_dioxide_concentrations_over_india/21978228-1-eng-GB/Nitrogen_dioxide_concentrations_over_India_pillars.jpg

*ESRL Global Monitoring Laboratory - Global Radiation and Aerosols*. (n.d.). Retrieved June 6, 2020, from https://www.esrl.noaa.gov/gmd/grad/surfrad/aod/

*European Space Agency*. (n.d.). Retrieved June 6, 2020, from http://www.esa.int/

*GES DISC Tools: Giovanni*. (n.d.). Retrieved June 6, 2020, from https://disc.gsfc.nasa.gov/information/tools?title=Giovanni

*Giovanni*. (n.d.). Retrieved June 6, 2020, from https://giovanni.gsfc.nasa.gov/giovanni/#service=TmAvMp&starttime=&endtime=

He, M. Z., Kinney, P. L., Li, T., Chen, C., Sun, Q., Ban, J., Wang, J., Liu, S., Goldsmith, J., & Kioumourtzoglou, M. A. (2020). Short- and intermediate-term exposure to NO2 and mortality: A multi-county analysis in China. *Environmental Pollution*, *261*(2), 114165. https://doi.org/10.1016/j.envpol.2020.114165

*In coronavirus lockdown extension, Modi wields stick, offers carrot on exit route - Coronavirus Outbreak News*. (n.d.). Retrieved May 9, 2020, from https://www.indiatoday.in/coronavirus-outbreak/story/in-coronavirus-lockdown-extension-modi-wields-stick-offers-carrot-on-exit-route-1666741-2020-04-14

*India, Day 1: World's Largest Coronavirus Lockdown Begins - The New York Times*. (n.d.). Retrieved June 6, 2020, from https://www.nytimes.com/2020/03/25/world/asia/india-lockdown-coronavirus.html

*India coronavirus: Millions under lockdown as major cities restrict daily life - CNN*. (n.d.). Retrieved June 6, 2020, from https://edition.cnn.com/2020/03/23/asia/coronavirus-covid-19-update-india-intl-hnk/index.html

*India dominates the list of the world's most polluted cities | World Economic Forum*. (n.d.). Retrieved June 6, 2020, from https://www.weforum.org/agenda/2020/03/6-of-the-world-s-10-most-polluted-cities-are-in-india/

*India Lockdown Extension News: Lockdown extended by 2 weeks, India split into red, green and orange zones - The Economic Times*. (n.d.). Retrieved June 6, 2020, from https://economictimes.indiatimes.com/news/politics-and-nation/govt-extends-lockdown-by-two-weeks-permits-considerable-relaxations-in-green-and-orange-zones/articleshow/75491935.cms

*INSAT-3D - ISRO*. (n.d.). Retrieved June 6, 2020, from https://www.isro.gov.in/Spacecraft/insat-3d

Jiang, F., Deng, L., Zhang, L., Cai, Y., Cheung, C. W., & Xia, Z. (2020). Review of the Clinical Characteristics of Coronavirus Disease 2019 (COVID-19). *Journal of General Internal Medicine*, *2019*. https://doi.org/10.1007/s11606-020-05762-w

Khoder, M. I. (2002). Atmospheric conversion of sulfur dioxide to particulate sulfate and nitrogen dioxide to particulate nitrate and gaseous nitric acid in an urban area. *Chemosphere*, *49*(6), 675–684. https://doi.org/10.1016/S0045-6535(02)00391-0

*MODIS Web*. (n.d.). Retrieved June 6, 2020, from https://modis.gsfc.nasa.gov/

*MoHFW | Home*. (n.d.). Retrieved June 6, 2020, from https://www.mohfw.gov.in/

Muhammad, S., Long, X., & Salman, M. (2020). COVID-19 pandemic and environmental pollution: A blessing in disguise? *Science of the Total Environment*, *728*, 138820. https://doi.org/https://www.who.int/docs/default-source/coronaviruse/situation-reports/20200506covid-19-sitrep-107.pd

PIB Delhi. (2020). Update on Novel Coronavirus: one positive case reported in Kerala. *Ministry of Health and Family Welfare*, 1601095. https://pib.gov.in/PressReleasePage.aspx?PRID=1601095

Porcheddu, R., Serra, C., Kelvin, D., Kelvin, N., & Rubino, S. (2020). Similarity in Case Fatality Rates (CFR) of COVID-19/SARS-COV-2 in Italy and China. *Journal of Infection in Developing Countries*, *14*(2), 125–128. https://doi.org/10.3855/jidc.12600

Rohde, R. A., & Muller, R. A. (2015). Air pollution in China: Mapping of concentrations and sources. *PLoS ONE*, *10*(8), 1–14. https://doi.org/10.1371/journal.pone.0135749

*Space based observation on changes in Air Quality during COVID-19 Lockdown period*. (n.d.).

*The Coronavirus Outbreak Is Curbing China's CO2 Emissions - IEEE Spectrum*. (n.d.). Retrieved June 6, 2020, from https://spectrum.ieee.org/energywise/energy/environment/coronavirus-outbreak-curbing-china-co2-emissions

*Thematic Strategy on air pollution (Text with EEA relevance)*. (n.d.).

WHO. (2020). *Coronavirus disease (COVID-19) Situation Report – 107*. https://www.who.int/docs/default-source/coronaviruse/situation-reports/20200506covid-19-sitrep-107.pdf

Wilder-Smith, A., & Freedman, D. O. (2020). Isolation, quarantine, social distancing and community containment: Pivotal role for old-style public health measures in the novel coronavirus (2019-nCoV) outbreak. *Journal of Travel Medicine*, *27*(2), 1–4. https://doi.org/10.1093/jtm/taaa020